\documentstyle[12pt,epsfig]{article}
\let\section=\subsection  \let\subsection=\subsubsection

\def\be{\begin{equation}}
\def\ee{\end{equation}}
\def\bea{\begin{eqnarray}}
\def\eea{\end{eqnarray}}
 
\def\br{\mbox{\boldmath $r$}}
\def\bm{\mbox{\boldmath $m$}}

\begin{document}
\begin{center}
{\large \bf Effective action for quantum Hall skyrmions.
}\\[5mm]
A. Teufel and H. Walliser \\[5mm]
{\small \it 
Fachbereich Physik, Universit\"at Siegen,  
D57068 Siegen, Germany} 
\end{center}

\begin{abstract}\noindent

Recently, an $O(3)$ type of effective action was presented
for the quantum Hall ferromagnet, which accounts for bag 
formation observed in microscopic Hartree-Fock calculations.
We apply this action in the soliton sector and
compare with Hartree-Fock results. We find good
agreement over the whole parameter range where skyrmions 
exist. The standard minimal $O(3)$ model
cannot explain the bag and has further shortcomings
connected with that fact.

\end{abstract} 

\bigskip
\leftline{PACS 73.43.-f, 73.21.-b, 12.39.Dc}  
\leftline{Keywords: skyrmions, $O(3)$-model}

\vspace{1cm}

Various experiments \cite{sepw95,ssplck00}
have revealed evidence that quantum Hall
ferromagnets exhibit charged excitations that are toplologically
stable spin--textures named skyrmions. In a recent revised
Hartree--Fock (HF) calculation \cite{w01}
based on the formulation by Fertig et. al. \cite{fbcd94,fbcdks97},
it was found that the magnetization
is not a unit vector, instead its length is substantially reduced 
in regions with non--zero excess charge (bag formation). 
This phenomenon is intuitively understood, because electrons
in quantum Hall systems are mobile such that in regions with
reduced electron density also the magnetization is expected
to be reduced. In this way, the charged spin--texture is intimately 
connected with a bag which accomodates the excess charge.

There has been a lot of effort to derive an effective field
theory for the quantum Hall ferromagnet 
\cite{f91,skkr93,mmygd95,bmv96,ab97,mm98,g99}, but all these
approaches restrict the magnetization to a unit vector field.
It was quite recently, that
an effective $O(3)$ type of effective action
was proposed in terms of a gradient expansion \cite{w01},
which actually accounts for bag formation.
With lengths measured in units of the magnetic length
$\ell = \sqrt{\hbar c /eB}$, and energies in units of the
Coulomb energy $e^2/\epsilon \ell$,
the full energy functional is given by 
\bea\label{action}
E[\hat{\bm}] &=& \frac{\rho_s}{2} \int d^2r 
\left[ \partial_i \hat{\bm} \partial_i \hat{\bm}
- \frac{1}{4} ( \partial_i \hat{\bm} \partial_i \hat{\bm} )^2 \right]
\nonumber \\
&+& \frac{1}{2} \int \int d^2r_1 d^2r_2 \rho^C(\br_1) V(\br_1-\br_2)
\rho^C(\br_2) \\
&+& \frac{g}{4\pi} \int d^2r \left[ 1 - \left( 1- 
\frac{1}{4} \partial_i \hat{\bm} \partial_i \hat{\bm} \right) 
\hat{m}_3 \right] + {\cal O} (6) \, ,
\nonumber
\eea
where $\rho_s$ represents the spin stiffness, e.g. 
$4 \pi \rho_s = \frac{1}{4} \sqrt{\pi/2}$ for the Coulomb interaction.
Then, for a given electron--electron interaction there is only
one dimensionless parameter in the theory, namely, the effective
Zeeman coupling $g=g_s \mu_B B/(e^2/\epsilon \ell)$.

In addition to the non--linear sigma, the interaction and Zeeman
terms of the standard minimal $O(3)$ 
model \cite{f91,skkr93,apfgd97,mm98,g99}, there
appears a symmetric fourth order term and a correction
to the Zeeman term. The most important amendment is that
the magnetization is no longer directly identified with the $O(3)$ 
unit vector field, instead it is related to it in a nontrivial
way
\be\label{magnet}
\bm = \sigma \cdot \hat{\bm} = \left( 1- 
\frac{1}{4} \partial_i \hat{\bm} \partial_i \hat{\bm} \right) 
\cdot \hat{\bm} + {\cal O} (4) \, .
\ee
This relation holds independently from the electron--electron 
interaction employed and
reflects the fact that the bag
is always present, for all soliton sizes. With the physical
charge density replaced by the topological density, 
\be
\rho^C = -\frac{1}{8\pi} \epsilon_{ij} 
\hat{\bm} ( \partial_i \hat{\bm} \times
\partial_j \hat{\bm} ) + {\cal O} (4)   \, ,
\label{charge}
\ee
it was argued in Ref. \cite{w01}, that the energy functional 
(\ref{action}) is unique up to
${\cal O} (4)$ in the gradient expansion and linear terms in the
Zeeman coupling $g$. For short ranged forces it was explicitely
shown that this energy functional gives the correct results for
extended spin--textures. In the following we will investigate the
case of the Coulomb interaction.

The classical equations of motion following from the energy functional
(\ref{action}) are solved by the hedgehog ansatz
\be\label{hedgehog}
\hat{\bm} =
\left( \begin{array}{c} \sin F(r) \cos \varphi \\
\sin F(r) \sin \varphi \\ \cos F(r) \end{array} \right) \, ,
\ee
where the angle function $F(r)$ determines the
orientation of the unit vector $O(3)$ field. With this ansatz
the azimuthal angle may be integrated, 
\bea \label{effhog}
E[F] & = & \frac{\rho_s}{2} \int d^2r 
\left[ F^{\prime 2} + \frac{\sin^2 F}{r^2}
- \frac{1}{4} \left( F^{\prime 2} + \frac{\sin^2 F}{r^2} \right)^2 
\right] \nonumber \\
& + & \frac{1}{\pi} \int d^2r_1 \, \rho^C (r_1) \int d^2r_2 \, 
\rho^C (r_2) \, \frac{1}{r_>} K \left(\frac{r_<}{r_>}\right)   \\
& + & \frac{g}{4\pi} \int d^2r
\left[ 1 - \cos F  
+ \frac{1}{4} \left( F^{\prime 2} + \frac{\sin^2 F}{r^2} \right) 
\cos F \right]  \, . \nonumber
\eea
Here, $K(x)$ represents a complete elliptic integral of the first
kind \cite{as65} with $r_>=\max(r_1,r_2)$ and $r_<=\min(r_1,r_2)$
respectively. The integro--differential equation for the
angle function is derived straightforwardly and solved with the
appropriate boundary conditions using familiar relaxation
methods \cite{apfgd97,mm98}.

In the following we compare results of (i) the microscopic
Hartree-Fock theory (HF), (ii) the effective field theory (EFT) defined
by the energy functional (\ref{action}) together with 
relation (\ref{magnet}), and (iii) the standard minimal
field theory (MFT).
In Fig.1 we show the dependence of the soliton energy 
on the Zeeman coupling. For extremely
small $g$, i.e. large solitons, all three curves approach the
Bogomol'nyi bound , $\frac{1}{4} \sqrt{\pi/2}$, of the 
famous Belavin--Polyakov (BP) soliton
\cite{bp75}. 
\begin{figure}[h]
\begin{center}
\epsfig{figure=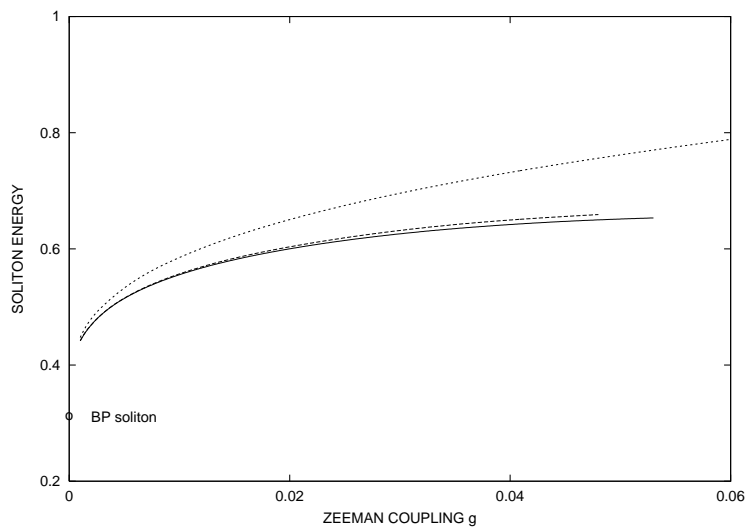,width=7cm,angle=270}
\protect\caption{Skyrmion energies for the Coulomb interaction 
as a function of the Zeeman coupling strength calculated with
HF (full line), EFT (dashed) and MFT (dotted). 
In the limit $g \to 0$ the Bogomol'nyi bound is approached. 
}
\end{center}
\end{figure}
With increasing $g$ then, the minimal $O(3)$
model starts to overestimate the soliton energy. In the HF
calculation there exists a critical Zeeman coupling 
$g_c \simeq 0.053$ beyond which solitons cease to exist. Although
we should not expect the gradient expansion to be reliable for too
small solitons, the energy functional (\ref{action}) reflects this
fact only at a somewhat smaller value $g_c \simeq 0.049$: the
attractive symmetric fourth order term is strengthend
and finally destroys the soliton.
In contrast, the minimal $O(3)$ model possesses stable soliton 
solutions for all Zeeman couplings.

Next, we are going to compare the profiles $F(r)$ and 
$\sigma(r)$, and the charge and spin densities
\bea
&&\sigma (r) =  1 - \frac{1}{4} \left( F^{\prime 2}(r) + 
\frac{\sin^2 F(r)}{r^2} \right) \, ,\\  
\label{density}
&&\rho^C (r) = -\frac{F^{\prime}(r) \sin F(r)}{4 \pi r} \, ,\\ 
&&m_3(r)  = \sigma (r) \cos F(r) 
\label{spindens}
\eea
for selected Zeeman couplings i.e. $g=0.01$ (larger soliton)
and $g=0.03$ (smaller soliton). As expected, 
larger deviations are observed of course for the smaller soliton,
where the gradient expansion converges more slowly.
\begin{figure}[h]
\begin{tabular}{cc}
\epsfig{figure=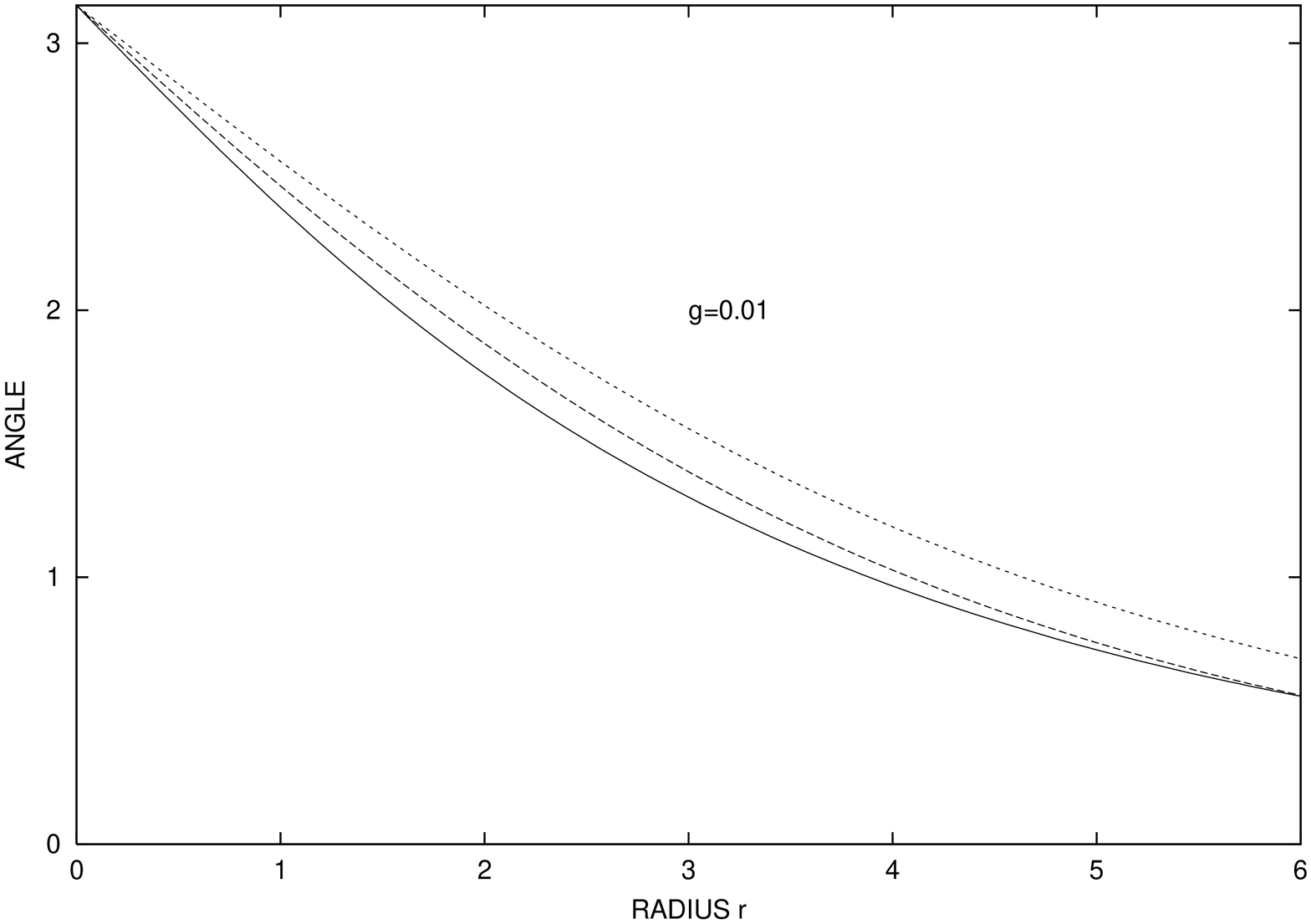,width=5cm,angle=270} 
&\epsfig{figure=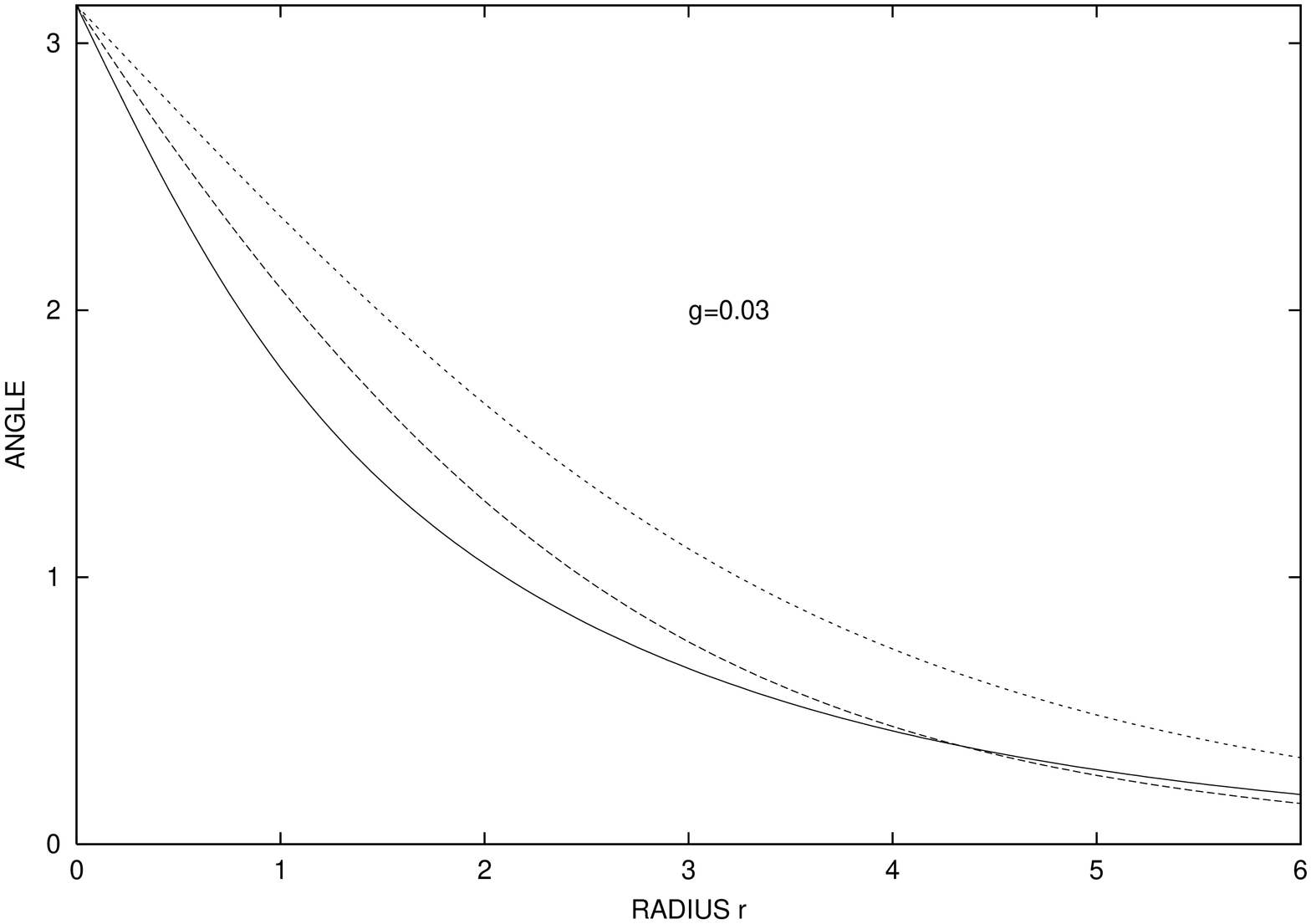,width=5cm,angle=270}
\end{tabular}
\protect\caption{Angle profiles for Zeeman couplings $g=0.01$
and $g=0.03$ versus the radius in magnetic lengths. Solid,
dashed and dotted lines refer to HF, EFT and MFT.
}
\begin{tabular}{cc}
\epsfig{figure=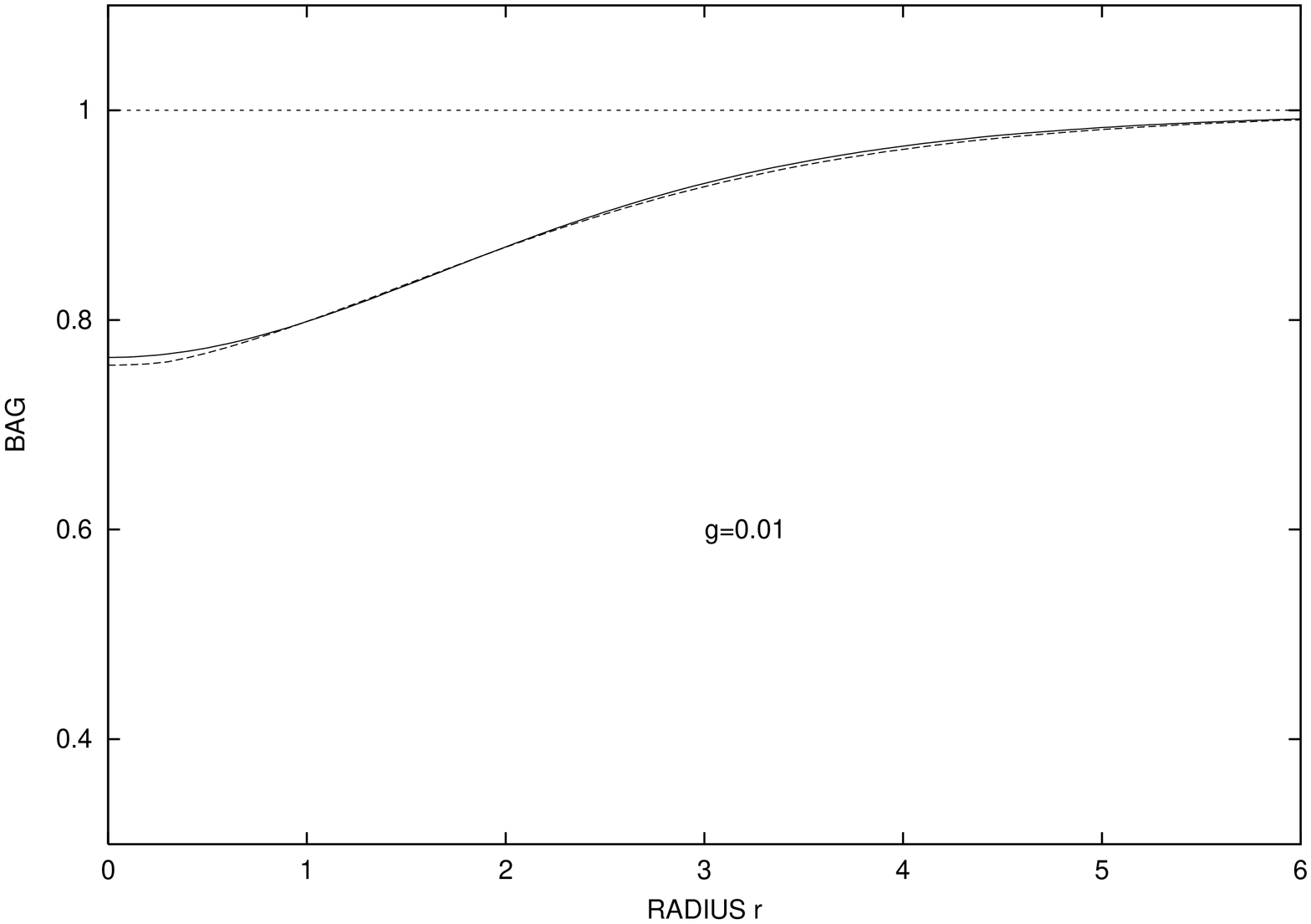,width=5cm,angle=270} 
&\epsfig{figure=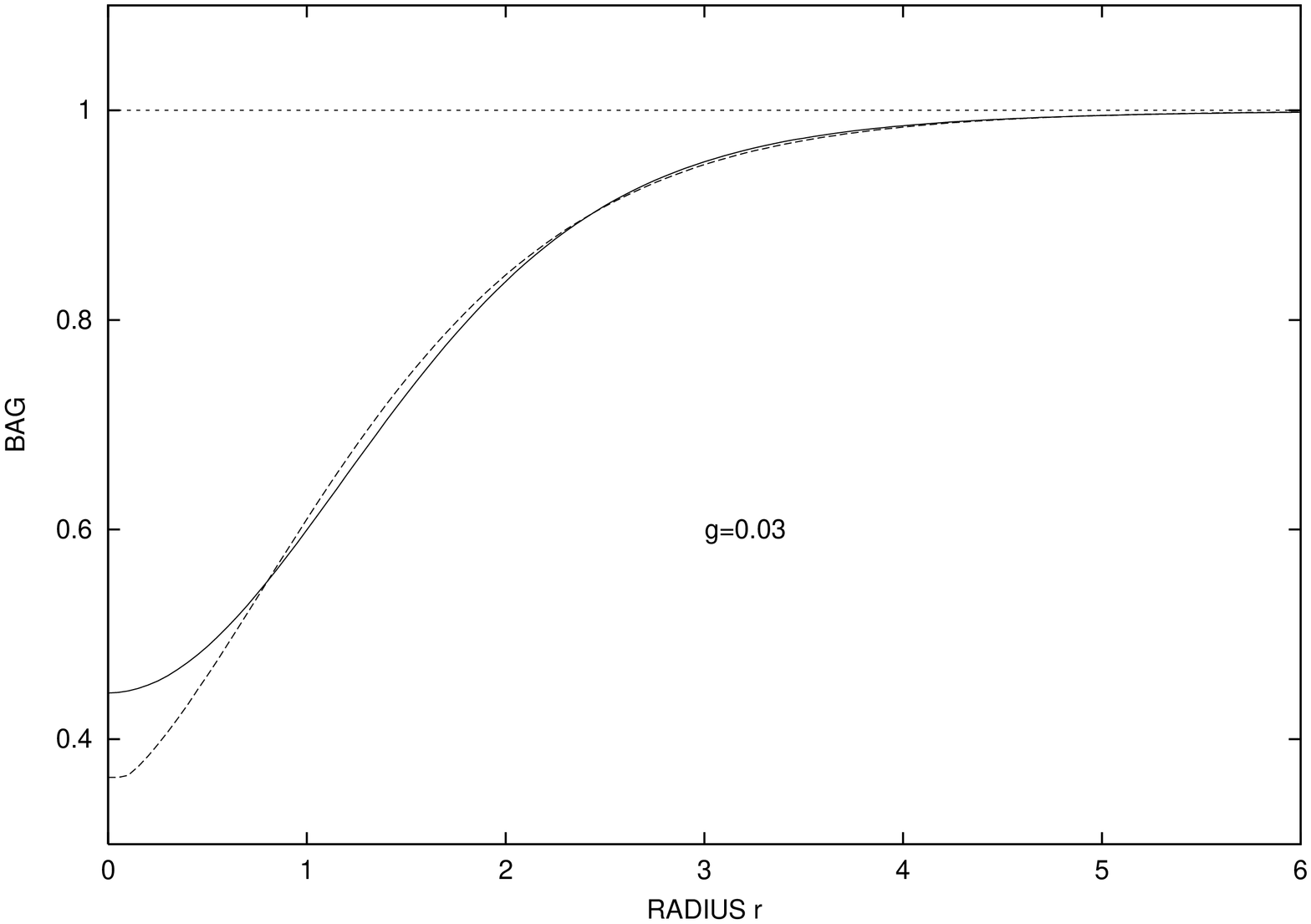,width=5cm,angle=270}
\end{tabular}
\protect\caption{Bag profiles for Zeeman couplings $g=0.01$
and $g=0.03$ versus the radius in magnetic lengths. Solid,
dashed and dotted lines refer to HF, EFT and MFT. There
exists no bag in the minimal field theory.
}
\end{figure}
Generally, the energy functional (\ref{action}) leads
to a much better agreement with the microscopic HF as compared to 
the minimal field theory. In particular, this improvement
is noticed for the bag shape functions depicted in Fig. 3:
there is no bag in the MFT and accordingly the corresponding profile
is $1$. From (\ref{spindens}) it follows then immediately
that the spin density starts always at $-1$ at the origin
(Fig. 5). For the spin densities of EFT and HF 
there exists no such restriction.
\begin{figure}
\begin{tabular}{cc}
\epsfig{figure=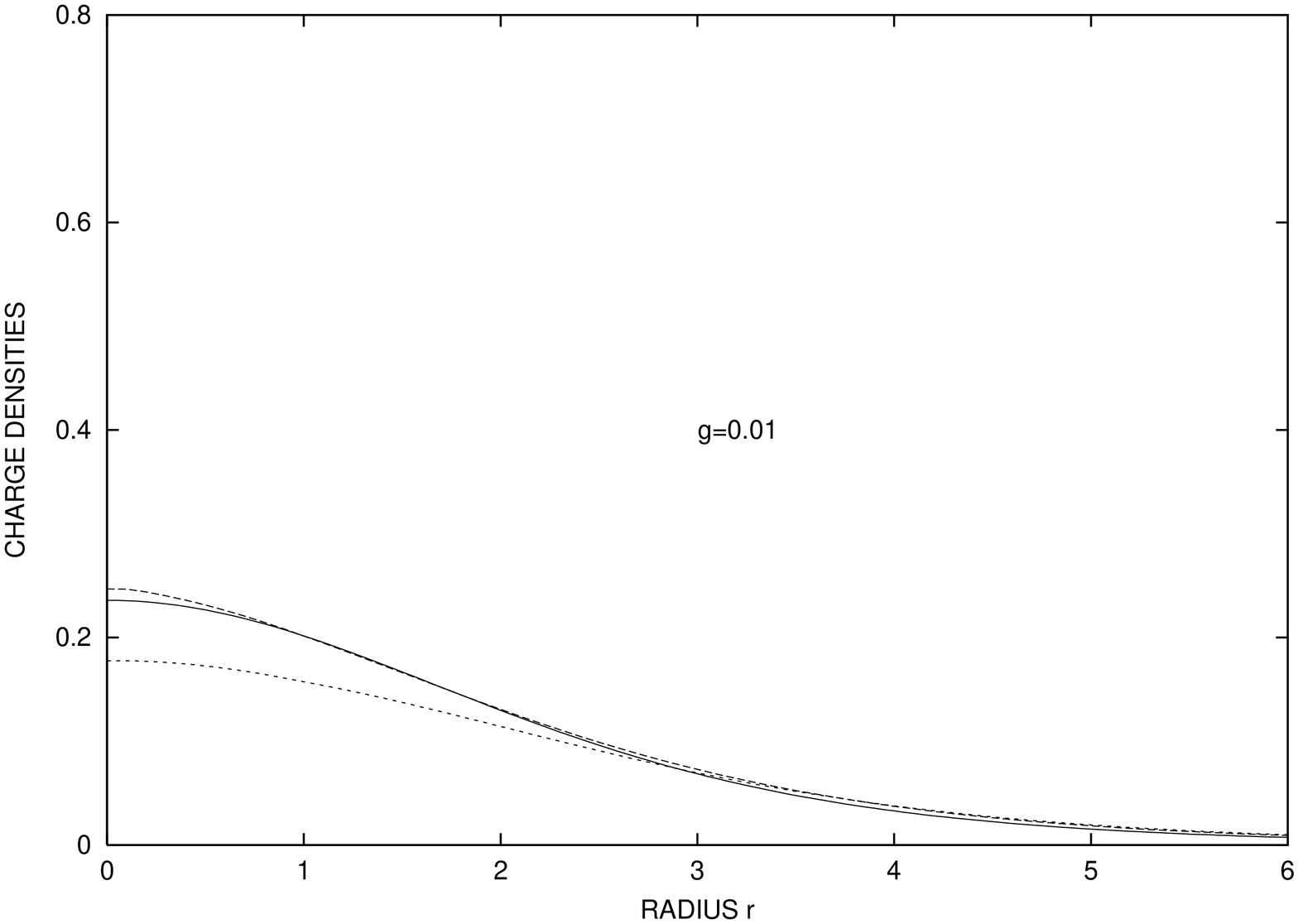,width=5cm,angle=270} 
&\epsfig{figure=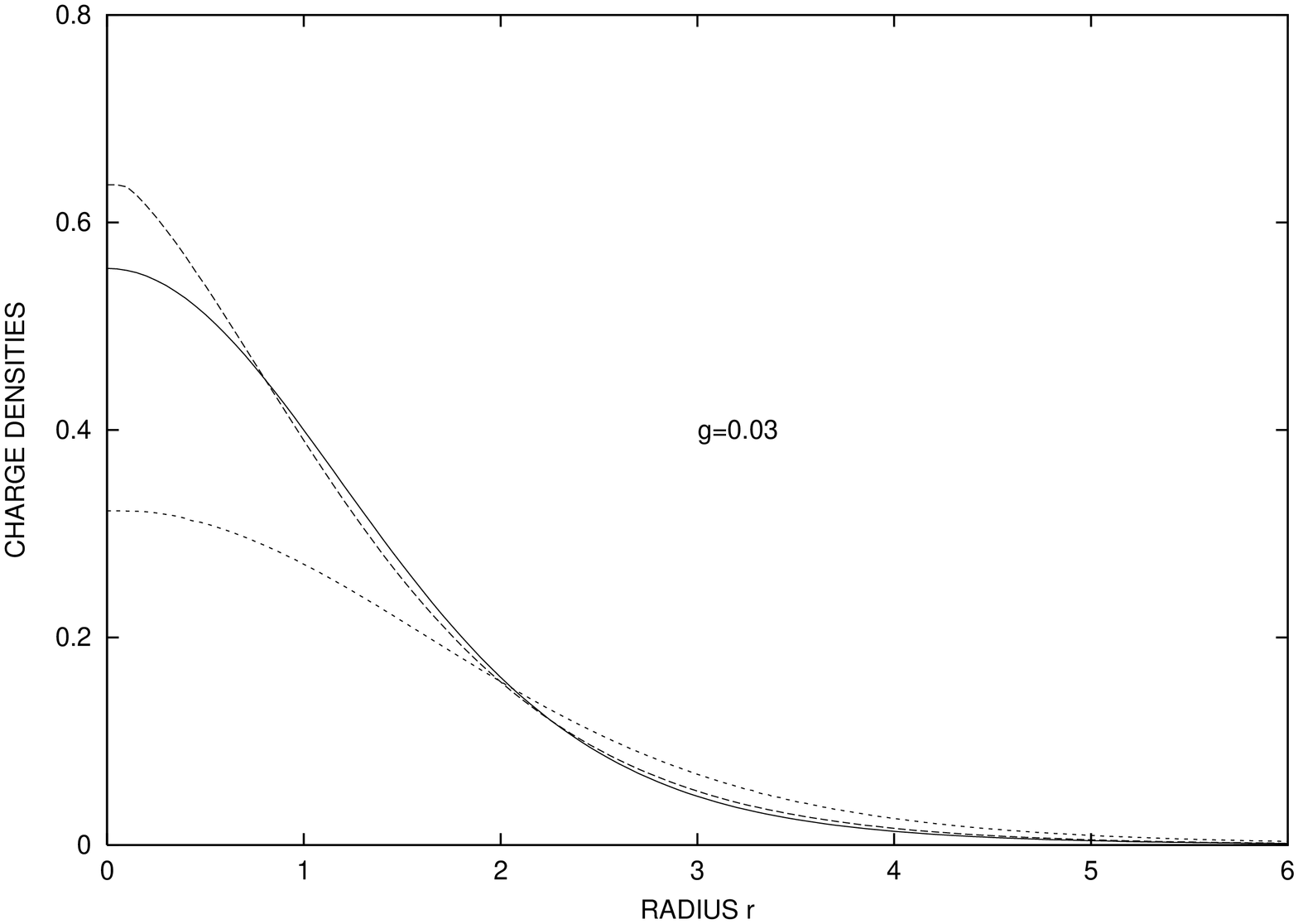,width=5cm,angle=270}
\end{tabular}
\protect\caption{Charge densities for Zeeman couplings $g=0.01$
and $g=0.03$ versus the radius in magnetic lengths. Solid,
dashed and dotted lines refer to HF, EFT and MFT.
}
\begin{tabular}{cc}
\epsfig{figure=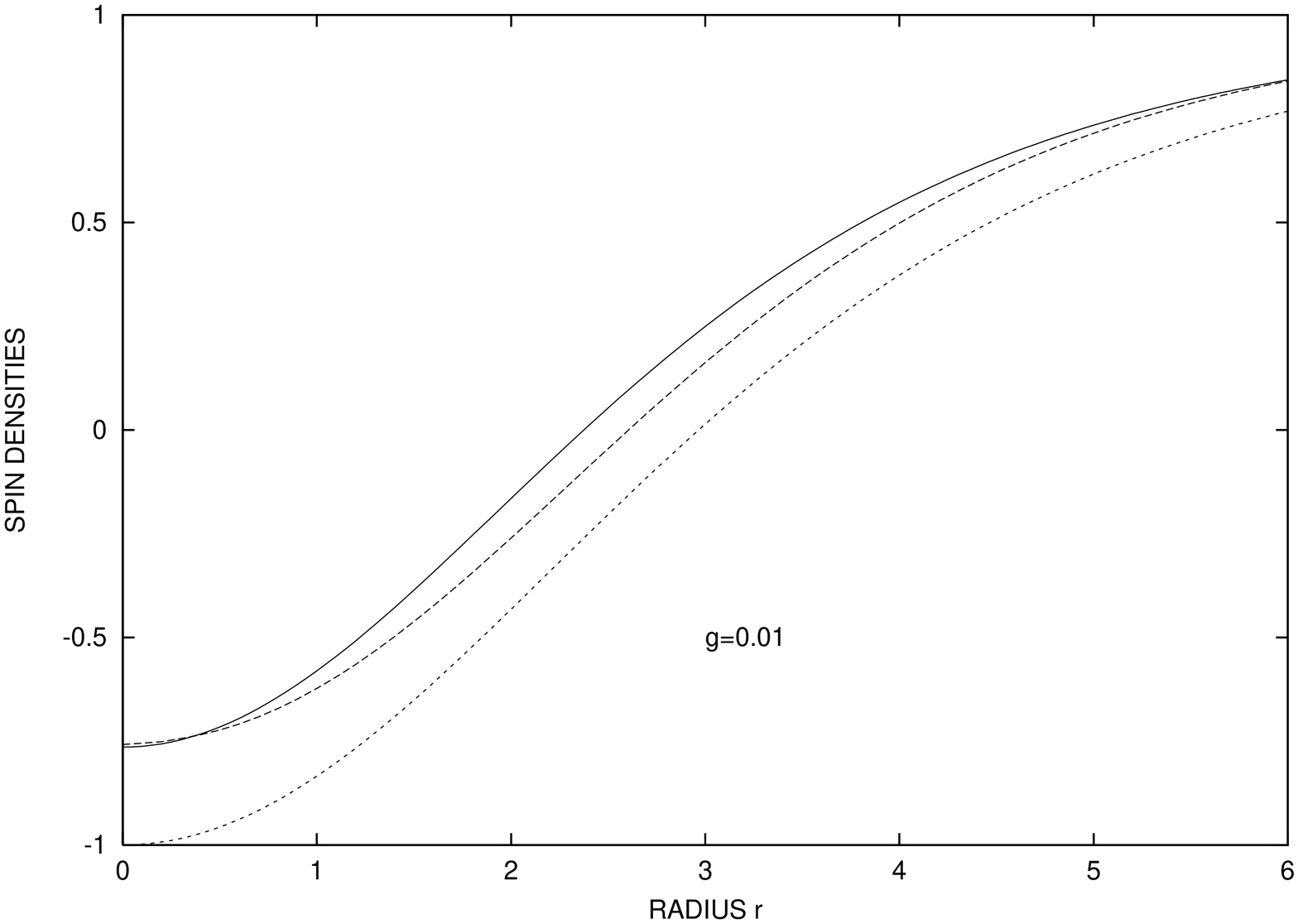,width=5cm,angle=270} 
&\epsfig{figure=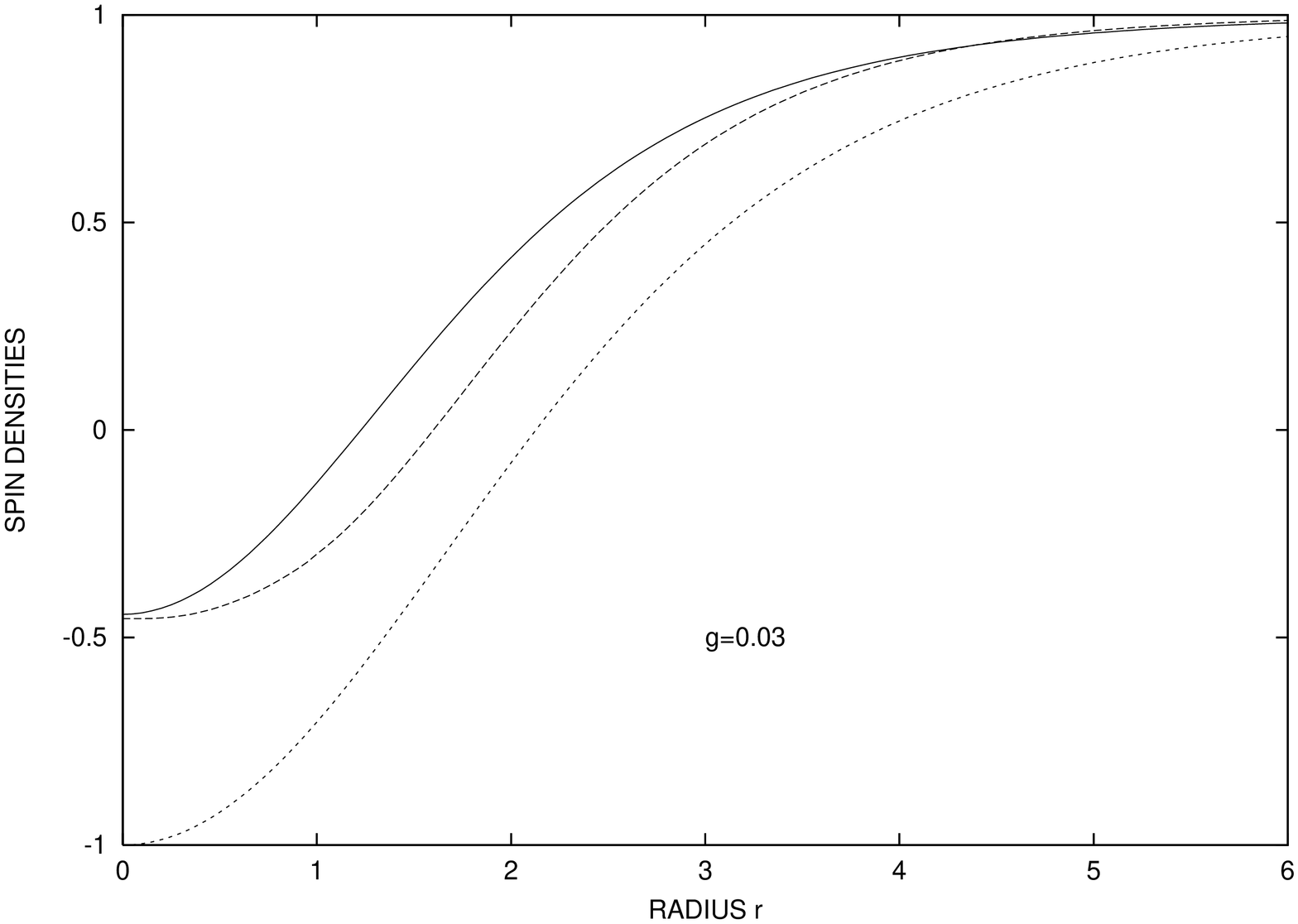,width=5cm,angle=270}
\end{tabular}
\protect\caption{Spin densities for Zeeman couplings $g=0.01$
and $g=0.03$ versus the radius in magnetic lengths. Solid,
dashed and dotted lines refer to HF, EFT and MFT. The spin 
density of the minimal field theory is confined to $-1$ at
the origin.
}
\end{figure}
Finally, we discuss the number of reversed spins defined as
the spatial integral over the spin density 
\be \label{spinflip}
K =  \frac{1}{4 \pi} \int d^2r \, (1 - m_3) 
- \frac{1}{2} \, .
\ee
The value $1/2$ is subtracted for convenience
such that the quasihole has $K=0$. The number of spin--flips
can be viewed as a measure for the soliton's size.
\begin{figure}[h]
\begin{center}
\epsfig{figure=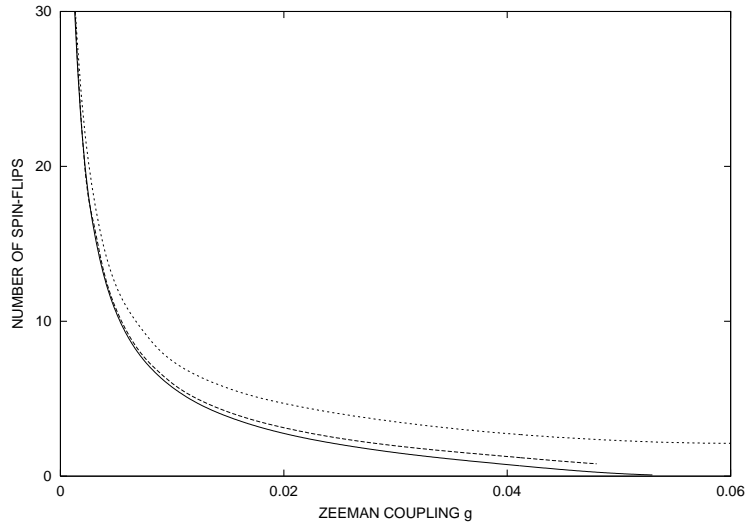,width=7cm,angle=270}
\protect\caption{Number of spin--flips as
a function of the Zeeman coupling strength calculated with
HF (full line), EFT (dashed) and MFT (dotted). 
}
\end{center}
\end{figure}
Its dependence on the Zeeman coupling (Fig. 6) 
behaves similar to that of the soliton energy (Fig. 1) 
for the three models. With increasing
coupling strength, the minimal $O(3)$ model starts to
overestimate the number of spin--flips. The EFT stays
close to the HF result practically over the whole region where
stable solitons exist. Since the Coulomb interaction
considered here represents the extreme case of a long ranged
force and the EFT becomes exact for short ranged forces, we  
may infer that this agreement is only improved when
more realistic interactions, which for example take the
finite thickness of the layer into account
\cite{fbcdks97,c97,mbo99}, are considered.


Concluding we may say that the energy functional (\ref{action})
gives a good description of skyrmions in the soliton sector.
The present investigation particularly supports our previous
statement that
the ${\cal O} (4)$ terms in the gradient expansion are unique.
Compared to the minimal $O(3)$ model this is considered a
considerable improvement. Through the skyrmion--skyrmion and
skyrmion--antiskyrmion interactions, the additional terms
in the effective action modify also the properties of multi--skyrmion
systems such as skyrmion gas and lattice \cite{cdbfgs97,nk98,mm99}.

Bag formation will also affect the 
time--derivative part of the effective action. This is
interesting and should be investigated.
Furthermore, it is desirable to derive an
effective field theory which takes the scalar field
as dynamical field seriously into account, without resorting
to the gradient expansion.

\end{document}